# Structural and transport properties of epitaxial $Na_xCoO_2$ thin films

A. Venimadhav

Department of Physics, The Pennsylvania State University, University Park, PA 16802

A. Soukiassian

Department of Materials Science and Engineering, The Pennsylvania State University, University Park, PA 16802

D. A. Tenne and Qi Li[a]

Department of Physics, The Pennsylvania State University, University Park, PA 16802

X. X. Xi

Department of Physics and Department of Materials Science and Engineering, The Pennsylvania State University, University Park, PA 16802

D. G. Schlom, R. Arroyave, and Z. K. Liu

Department of Materials Science and Engineering, The Pennsylvania State University, University Park, PA 16802

H. P. Sun and Xiaoqing Pan

Department of Materials Science and Engineering, The University of Michigan, Ann Arbor, MI 48109

Minhyea Lee and N. P. Ong

Department of Physics, Princeton University, NJ 08544

---

[a] Electronic mail: qil1@psu.edu

**Abstract**

We have studied structural and transport properties of epitaxial $Na_xCoO_2$ thin films on (0001) sapphire substrate prepared by topotaxially converting an epitaxial $Co_3O_4$ film to $Na_xCoO_2$ with annealing in Na vapor. The films are *c*-axis oriented and in-plane aligned with $[10\bar{1}0]$ $Na_xCoO_2$ rotated by 30º from $[10\bar{1}0]$ sapphire. Different Na vapor pressures during the annealing resulted in films with different Na concentrations, which showed distinct transport properties.

Layered cobaltate $Na_xCoO_2$ has attracted much attention recently due to its exceptional properties.[1] It has an unusually high thermoelectric power with low mobility, low resistivity, and high carrier density.[1] The Fermi surface [2] and electrical properties [3] of $Na_xCoO_2$ depend on the Na concentration: $Na_xCoO_2 \cdot 1.3H_2O$ is a superconductor for $x$ around 0.3;[4,5] at $x = 0.5$ it is a charge-ordered insulator;[6] and at higher Na concentrations it becomes a metal following the Curie-Weiss law.[3,7] The triangular structure of the $CoO_2$ planes and the strong electron correlation effect have been recognized as sources of rich properties of $Na_xCoO_2$.[8] For example, the large thermopower in $Na_xCoO_2$ has been attributed to the spin entropy due to the strong electron correlation effects.[9] $Na_xCoO_2$ has been prepared in polycrystalline and single crystalline forms, but there are very few reports on $Na_xCoO_2$ thin films.[10-12] Recently, Ohta *et al.* [10] reported epitaxial $Na_xCoO_2$ films by reactive solid phase epitaxy, however the Na concentration in the films was not well controlled. In this paper we describe the structural and transport properties of epitaxial $Na_xCoO_2$ thin films fabricated by a process which is similar to that of Ohta *et al.* [11], but allows some degrees of control of Na concentration in the film. Films with different Na concentrations showed very different transport properties.

The epitaxial $Na_xCoO_2$ films were fabricated using a two-step process. First, an epitaxial $Co_3O_4$ film was grown by pulsed laser deposition (PLD) on a (0001) sapphire substrate. A KrF excimer laser was used with an energy density of 3.7 J/cm$^2$ on a CoO target. The substrate was kept at 650-700°C during the deposition in 200 mTorr flowing oxygen. At a repetition rate of 8 Hz, the deposition rate is 0.11 Å/s. The $Co_3O_4$ film was then sealed in an alumina crucible with sodium bicarbonate ($NaHCO_3$) or sodium acetate (NaOOCCH$_3$) powder and heated to 800ºC for 2.5 hours to form the Na$_x$CoO$_2$ film. A topotaxial conversion occurred, during which the crystallographic alignment of Co$_3$O$_4$ was inherited by Na$_x$CoO$_2$. The thickness of the Co$_3$O$_4$ film was around 1600 Å, which became ~3000 Å following the topotaxial conversion to Na$_x$CoO$_2$.

X-ray diffraction scans of an epitaxial Co$_3$O$_4$ film on a (0001) sapphire substrate are shown in Fig. 1. Co$_3$O$_4$ has a spinel structure with a space group *Fd3m*. The $\theta$-$2\theta$ scan in Fig. 1(a) shows only peaks arising from diffraction off (111) Co$_3$O$_4$ planes apart from the substrate peak, indicating a phase-pure Co$_3$O$_4$ film with [111] direction normal to the substrate surface. The rocking curve of the Co$_3$O$_4$ 111 peak had a full width at half maximum (FWHM) of 0.24° in $\omega$, equal to our instrumental resolution. A lattice parameter $a$ = 8.087 ± 0.001 Å was obtained. A $\phi$ scan of the 220 Co$_3$O$_4$ peak is shown in Fig. 1(b), where $\phi$ = 0° is aligned parallel to the [10$\bar{1}$0] in-plane direction of the sapphire substrate. The presence of six 220 peaks (where a single crystal would show only three) indicates an epitaxial Co$_3$O$_4$ film with two twinned variants related by a 60° rotation. The FWHM in $\phi$ is 0.55°. The in-plane epitaxial relationship is that [110] Co$_3$O$_4$ is rotated by ±30° from [10$\bar{1}$0] Al$_2$O$_3$.

X-ray diffraction scans of a Na$_x$CoO$_2$ film, which was converted from a Co$_3$O$_4$ film by annealing with NaHCO$_3$ powder, are shown in Fig. 2. Na$_x$CoO$_2$ has a hexagonal *P6$_3$22* structure. In the $\theta$-$2\theta$ scan in Fig. 2(a), 00$l$ peaks of Na$_x$CoO$_2$ are observed besides a substrate peak, indicating a *c*-axis oriented film. A weak peak of NaHCO$_3$ is also present due to the NaHCO$_3$ dust on the film surface resulting from the annealing process, which is also confirmed by a Raman scattering measurement. The $\phi$-scan of the 10$\bar{1}$2 Na$_x$CoO$_2$ peak is shown in Fig. 2(b), where $\phi$ = 0 is parallel to the [10$\bar{1}$0] direction of the sapphire substrate. The six-fold symmetry indicates a topotaxial conversion from the epitaxial $Co_3O_4$ film with the angle between $[10\bar{1}0]$ $Na_xCoO_2$ and $[10\bar{1}0]$ sapphire being 30°. Lattice constants $c = 11.02 \pm 0.003$ Å and $a = 2.456 \pm 0.003$ Å were obtained. The rocking curves showed broad FWHM of 2° in $\omega$ and 1.02° in $\phi$. These values indicate that the crystalline quality of the topotaxially converted $Na_xCoO_2$ film is not as high as that of the starting $Co_3O_4$ film.

Figure 3(a) is a bright field cross-sectional transmission electron microscopy (TEM) image of a $Na_xCoO_2$ film. It shows the film with a smooth surface. A white line is seen in the middle of the film corresponding to a thin layer of amorphous material. An x-ray energy dispersive spectroscopy (EDS) analysis shows that the Na concentration in such amorphous layers is higher than that in the crystalline $Na_xCoO_2$ film. A much thicker amorphous layer was observed at the film/substrate interface, whose chemical composition is similar to the substrate ($Al_2O_3$). Figure 3(b) and 3(c) are selected area electron diffraction (SAED) patterns corresponding to the film and the substrate, respectively. They show an epitaxial relationship between the $Na_xCoO_2$ film and the substrate of $Na_xCoO_2$ (0001)$[10\bar{1}0]$ ∥ sapphire (0001)$[2\bar{1}\bar{1}0]$, which is consistent with the x-ray diffraction analysis. The smeared intensity distribution of reflections in Fig. 3(b) indicates distortions of crystal planes in the thin film, consistent with high-resolution TEM observations of waviness in the $Na_xCoO_2$ lattice planes. Details of the microstructure investigation of the $Na_xCoO_2$ films will be published elsewhere.

The Na concentration of the $Na_xCoO_2$ films depends on the powder used for the annealing. At 800ºC, the equilibrium vapor pressure is 0.155 Torr for $NaHCO_3$ and 444 Torr for $NaOOCCH_3$. EDS measurements show that the Na concentration is $x = 0.68$

±0.03 for films annealed in $NaHCO_3$ and $x$ = 0.75 ±0.02 for films annealed in $NaOOCCH_3$. Figure 4 shows the resistivity versus temperature curves of two $Na_xCoO_2$ films with different Na concentrations. The temperature dependence for the film annealed in $NaOOCCH_3$, marked by "$x$ = 0.75", is characteristic of bulk and single crystal $Na_xCoO_2$ samples with $x$ = 0.75.[3,12] The downturn at low temperatures has been attributed to a phase transition to an antiferromagnetic spin-density-wave.[13] The resistivity behavior of the film annealed in $NaHCO_3$, marked by "$x$ = 0.68", is consistent with single crystals with lower Na concentrations.[3] The inset to Fig. 4 shows the thermopower, $S$, versus temperature for a film with $x$ = 0.68. A temperature gradient was generated by a resistive heater attached on one end of the film while the other end was mounted on a cold finger. A pair of type-$E$ (chrome-constantan) thermocouples and a pair of 25 μm gold wires were used to measure the temperature gradient and the difference of electric potential, respectively, to obtain the thermoelectirc power. The magnitude of $S$ at 300 K as well as the overall temperature dependence shown in the figure are consistent with the result from the in-plane measurement of single crystals with $x$= 0.7. [9] These results further confirm the Na concentration measurements by EDS.

In conclusion, epitaxial thin films of $Na_xCoO_2$ were prepared by annealing PLD grown epitaxial $Co_3O_4$ films in Na vapor. The Na compounds used during annealing, $NaHCO_3$ and $NaOOCCH_3$, have different Na vapor pressures, resulting in $Na_xCoO_2$ films of two different Na concentrations. The topotaxial conversion led to a poorer crystallinity in the $Na_xCoO_2$ films than in $Co_3O_4$ films. Nevertheless, the films are $c$-axis oriented with in-plane alignment with the substrate. The temperature dependent transport properties are distinctly different for films of different Na concentrations, and they are consistent with the bulk results. Our results demonstrate that some degrees of control of the Na concentration in the $Na_xCoO_2$ films can be achieved by using Na compounds of different vapor pressures during annealing.

The work was partially supported by NSF under Grant Nos. DMR-0405502 (Li), DMR-0103354 (Xi, Schlom), DMR-0205232 (Liu), and DMR-0308012 (Pan), by DOE under Grant Nos. DE-FG02-01ER45907 (Xi) and DE-FG02-97ER45638 (Schlom), and by ONR under Grant No. N00014-04-1-0057 (Ong).

FIGURE CAPTIONS

Figure 1. (a) $\theta$-$2\theta$ x-ray diffraction scan of a $Co_3O_4$ film grown on a (0001) sapphire substrate. The 0006 sapphire substrate peak is marked by an asterisk (*). (b) $\phi$-scan of the 220 $Co_3O_4$ peak at $\chi = 54.7°$, indicating that the film is epitaxial. $\phi = 0$ is parallel to the $[10\bar{1}0]$ in-plane direction of the substrate.

Figure 2. (a) $\theta$-$2\theta$ x-ray diffraction scan of a $Na_xCoO_2$ film on a (0001) sapphire substrate. The 0006 sapphire substrate peak is marked by an asterisk (*). (b) $\phi$-scan of the $10\bar{1}2$ $Na_xCoO_2$ peak at $\chi = 23.6°$, indicating that the film is epitaxial. $\phi = 0$ is parallel to the $[10\bar{1}0]$ in-plane direction of the substrate.

Figure 3. (a) Bright field TEM image of a $Na_xCoO_2$ film on a sapphire substrate. The white line in the middle of the film corresponds to a thin layer of Na-rich amorphous material. (b) SAED pattern from the film. (c) SAED pattern corresponding to the substrate.

Figure 4. Resistivity versus temperature curves for two $Na_xCoO_2$ films with different Na concentrations. Inset: Thermopower versus temperature for a $Na_xCoO_2$ film with $x=0.68$.

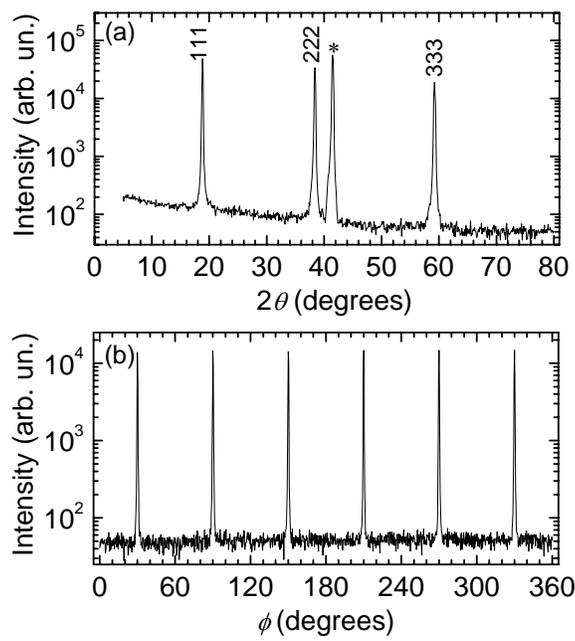

Fig. 1 of 4

Venimadhav et al.

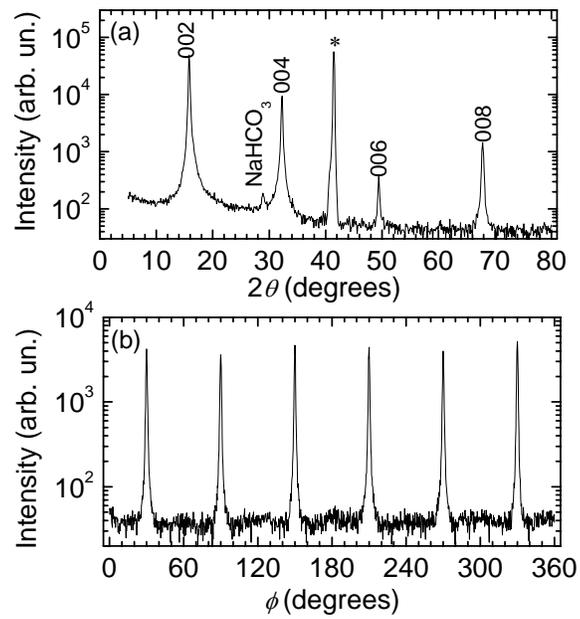

Fig. 2 of 4

Venimadhav et al.

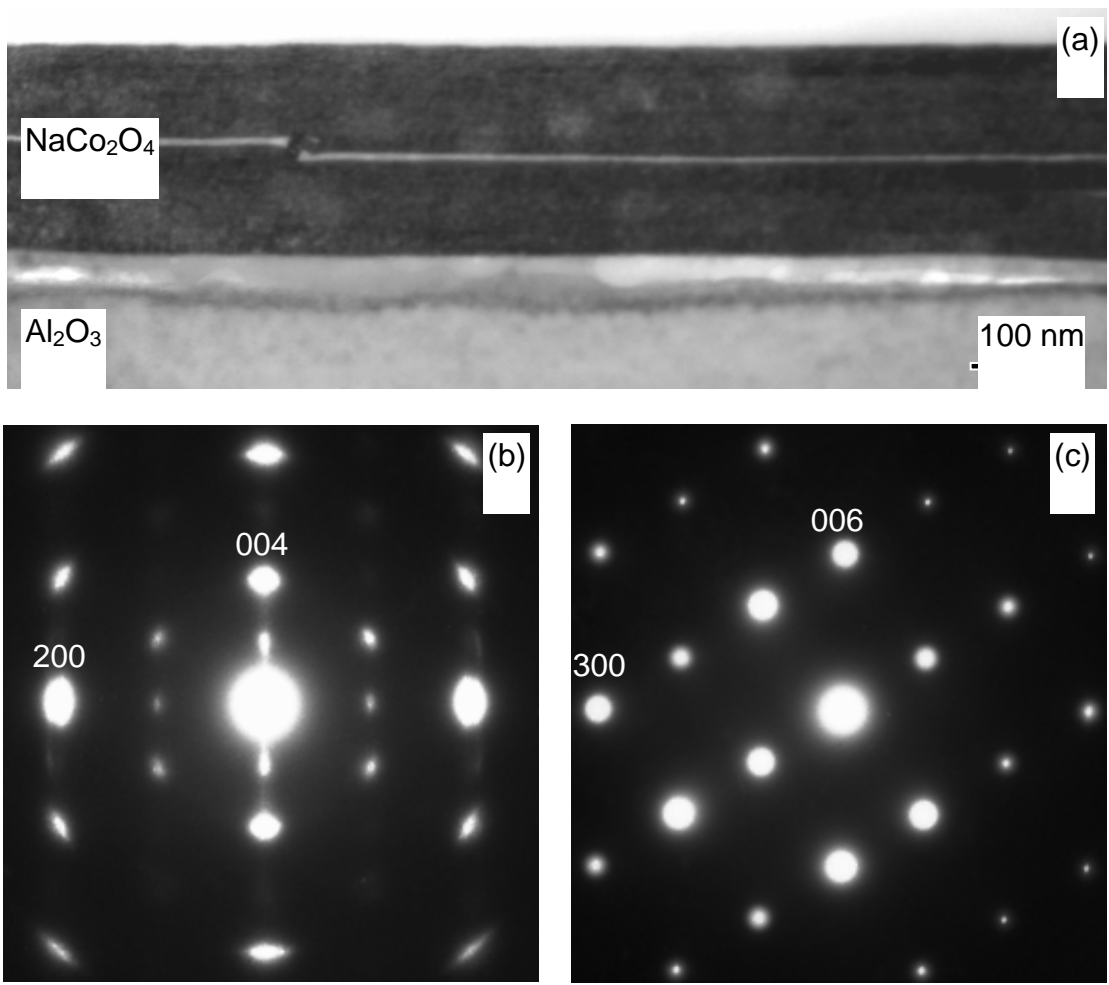

Fig. 3 of 4

Venimadhav et al.

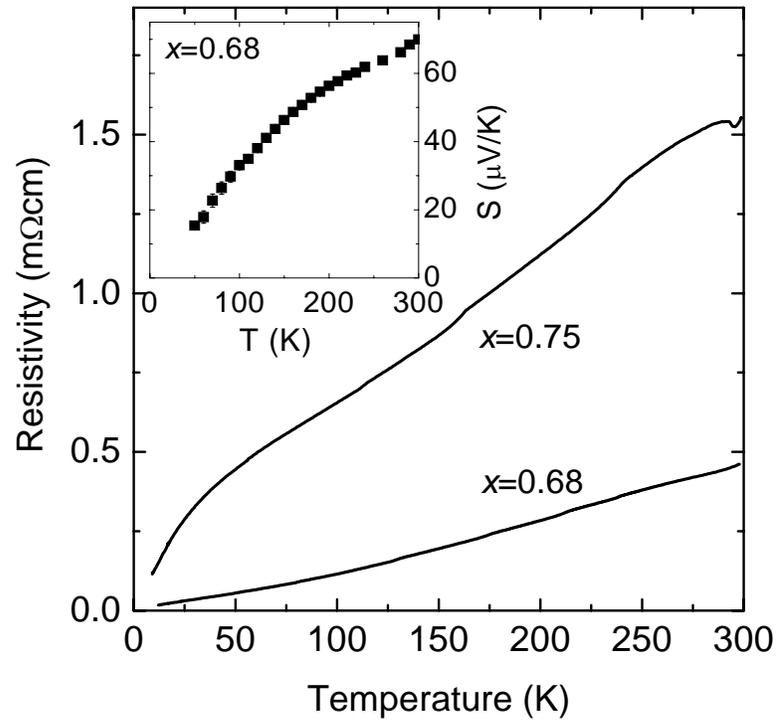